\documentclass[fleqn,usenatbib]{mnras}

\usepackage{newtxtext,newtxmath}

\usepackage[T1]{fontenc}
\makeatletter
\newlength{\abovecaptionskip}%
\makeatother
\usepackage{threeparttable}%

\DeclareRobustCommand{\VAN}[3]{#2}
\let\VANthebibliography\thebibliography
\def\thebibliography{\DeclareRobustCommand{\VAN}[3]{##3}\VANthebibliography}

\usepackage{graphicx}	
\usepackage{amsmath}	

\title[]{Period-Colour and Amplitude-Colour relations for OGLE-$\delta$~Scuti stars in the Galactic Bulge and LMC}

\author[M Deka et al.]{
Mami Deka,$^{1}$\thanks{E-mail:mamideka8@gmail.com}
Shashi M. Kanbur,$^{2}$\thanks{E-mail:shashi.kanbur@oswego.edu }
Sukanta Deb,$^{1,3}$\thanks{E-mail: sukanta.deb@cottonuniversity.ac.in}
Susmita Das,$^{4}$
Kerdaris Kurbah,$^{1}$
Earl P. Bellinger,$^{5,6}$
\newauthor{Anupam Bhardwaj$^{7}$}
\\
$^{1}$Department of Physics, Cotton University, Panbazar, Guwahati 781001,Assam, India\\
$^{2}$Department of Physics, State University of New York Oswego, Oswego, NY 13126, USA\\
$^{3}$Space and Astronomy Research Center, Cotton University, Panbazar, Guwahati 781001, Assam, India\\
$^{4}$Konkoly Observatory, CSFK, Konkoly Thege Miklós út 15-17, H-1121, Budapest, Hungary\\
$^{5}$Max Planck Institute for Astrophysics, Garching, Germany\\
$^{6}$Stellar Astrophysics Centre, Aarhus, Denmark \\
$^{7}$INAF-Osservatorio Astronomico di Capodimonte, Via Moiariello 16, I-801301, Napoli, Italy
}
\date{Accepted XXX. Received YYY; in original form ZZZ}

\pubyear{2022}

\begin{document}
\label{firstpage}
\pagerange{\pageref{firstpage}--\pageref{lastpage}}
\maketitle

\begin{abstract}
We present an analysis on the behaviour of the Galactic bulge and the Large Magellanic Cloud (LMC) $\delta$~Scuti stars in terms of period-colour and amplitude-colour (PCAC) relations at maximum, mean and minimum light.  The publicly available Optical Gravitational Lensing Experiment-IV (OGLE-IV) light curves for Galactic bulge and OGLE-III light curves for LMC $\delta$~Scuti stars are exploited for the analysis. It has been found that the Galactic bulge $\delta$~Scuti stars obey flat PC relations at \textit{maximum/mean/minimum} light while the LMC $\delta$~Scutis have \textit{sloped/sloped/flat} PC relations at \textit{maximum/mean/minimum} light.  Both the Galactic bulge and the LMC $\delta$~Scutis have \textit{sloped/flat/sloped} AC relations at \textit{maximum/mean/minimum}. These relations also show that Galactic $\delta$~Scutis are hotter as compared to their LMC counterparts. The period-amplitude (PA) relations for $\delta$~Scutis exhibit different behaviour in the Galactic bulge and the LMC. The LMC variables are found to have higher amplitudes at a given period. The amplitude of the Galactic bulge $\delta$ Scuti shows a bimodal distribution which can be modelled using a two-component Gaussian Mixture Model: one component with a lower amplitude and another with a higher amplitude. The observed behaviour of the $\delta$~Scuti PCAC relations can be explained using the theory of the interaction of hydrogen ionization front (HIF) and stellar photosphere as well as the PA diagram. We use MESA-RSP to calculate theoretical non-linear hydrodynamical pulsation models for $\delta$~Scuti stars with input metallicities of $Z=0.02$ and $Z=0.008$ appropriate for the Galactic bulge  and LMC, respectively. The observed PCAC relations and theoretical calculations support the HIF-photosphere interaction theory.
\end{abstract}
\begin{keywords}
stars: evolution-stars:variables:Delta Scuti-Galaxy:bulge-Magellanic Clouds 
\end{keywords}



\section{Introduction}
$\delta$~Scutis are pulsating variable and intermediate mass stars with masses in the range $(1.0-3.0)~{M}_{\odot}$ having spectral types between A2 and F2 located at the 
intersection of the Cepheid instability strip with the main sequence \citep{goup05,jaya20,bedd20}. Their pulsation periods lie in the range $(0.03-0.3)$ days \citep{breg79}.  They exhibit
a wide range of metallicity and temperature;  and pulsate in both single- and  multi-modes \citep{temp00,murp19}. They serve as an important astrophysical tool to test the theories of stellar pulsation and evolution. The single-mode $\delta$~Scuti obey a period-luminosity (PL) relation, which make them reliable distance indicators \citep{mcna97,mcna11,ziaa19,poro21}, while multi-mode $\delta$~Scuti stars can be used to understand the properties of deep stellar interiors \citep{breg98}.

$\delta$~Scuti stars can be divided into various subclasses based on their metallicities, light curve shapes and amplitudes. On the basis of metallicity, they can be classified into  Population-I $\delta$~Scuti stars (metal rich), and Population-II SX~Phoenicis stars (metal poor). However, not all the SX~Phoenicis are found to be metal poor \citep{neme17}. 
A clear distinction between these two populations of  $\delta$~Scuti stars is yet not possible and they may therefore be considered as stars having mixed populations \citep{neme17,guzi21}.
Further, based on their pulsation behaviour, they are divided into two classes: high amplitude $\delta$~Scuti stars (HADS-) and low amplitude $\delta$~Scuti stars (LADS). HADS can be found in the post main-sequence region of the instability strip, while low amplitude 
$\delta$~Scuti stars are located in all regions of the $\delta$~Scuti instability strip \citep{temp00,chan13}.

The period-colour and amplitude-colour (PCAC) relations have been used extensively to study the radiation hydrodynamics of outer envelope structure and evolutionary status of Cepheids and RR~Lyraes \citep{simo93,kanb04,bhar14,ngeo17,das18}. These studies have found that the PCAC relations of different types of pulsating stars show different
behaviour at maximum/mean/minimum light; long period $(P>10$~days) Classical Cepheids exhibit flat/sloped PC relations at maximum/minimum light, while RR~Lyraes show sloped/flat PC relations at maximum/minimum light 
\citep[and references therein]{kanb04,bhar14}. \citet{das20} extended this further to include Type~II Cepheids. They found that the contrasting behaviour of the PCAC relations are strongly correlated with the relative location of HIF and stellar photosphere. The photosphere is considered being at optical depth $\tau=\frac{2}{3}$, and the HIF is defined to be the region where the majority of hydrogen becomes ionized. The stellar photosphere and HIF are not always co-moving during a pulsation cycle. The HIF moves ``in and out'' within the mass distribution during a pulsation cycle. Besides, the relative location of the HIF and the stellar photosphere is pulsation-phase dependent. The HIF interacts with the photosphere at those phases where the photosphere lies at the base of HIF. This has been well-established in the literature \citep{simo93,kanb95,kanb96,kanb04,bhar14,ngeo17,das18,das20}.

The PCAC relations as a function of phase are important probes of the structure of the outer envelope and offer an insight into the physics of stellar pulsation and evolution \citep{simo93,kanb95}. They also influence the period-luminosity (PL) relations which is crucial for the distance scale  and non-cosmic microwave background estimates of Hubble's constant \citep{beat16,reis16,reis19}. Changes in the behaviour of the PC relation can be reflected on the PL relation through the PLC relation as PC/PL relations are just the projection of the PLC (period-luminosity-colour) relation on either the PC or PL (period-magnitude) planes. Since the PLC/PL/PC relations at mean light are the average of the corresponding relations through pulsation phase, changes in these relations at a particular phase or range of phases can be reflected in changes in the relations at mean light. For example, \citet[and references therein]{das20} showed evidence of the nonlinear PC relation for LMC Cepheids at certain phases. This leads to 
a nonlinear PL relation at those phases \citep{bhar16}. This effect is reduced at mean light. The theory initiated in \citet*{simo93} and developed in \citet[and references therein]{das20} demonstrate how the interaction  of the stellar photosphere and HIF can produce such changes in the multiphase PC relation.

In this study, 
we have investigated the observed PCAC relations at maximum/mean/minimum light of $\delta$~Scuti stars 
for the first time and verifies the HIF-photosphere interaction theory of \citet*{simo93} using 
MESA-RSP code. 

The remaining paper is organized as follows: In Section~\ref{sec:data}, the 
selection and cleaning criteria of data is discussed. The methods and analysis criteria are discussed in Section~\ref{sec:Methods}. Section~\ref{sec:Results} describes the results of this work. Finally, we 
summarize the results of the work in Section~\ref{sec:Summary}.

\begin{figure*}
\includegraphics[width=1.0\textwidth,keepaspectratio]{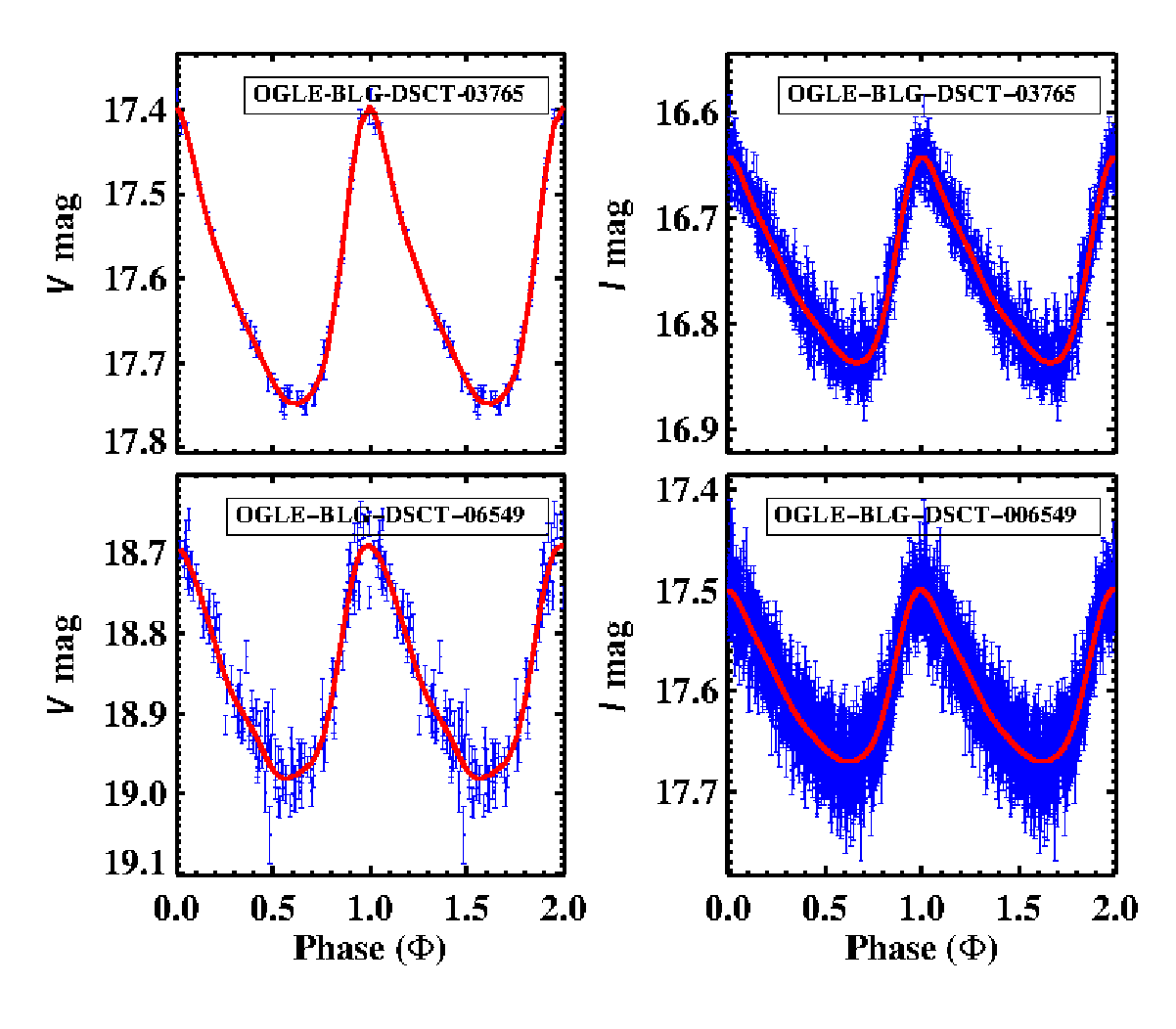}
\caption{Two randomly selected light curves of Galactic bulge 
$\delta$~Scuti stars in $V$- and $I$- band. The red solid line represents the Fourier 
fitted curve.}
\label{fig:lc_blg}
\end{figure*}
\begin{figure*}
\includegraphics[width=1.0\textwidth,keepaspectratio]{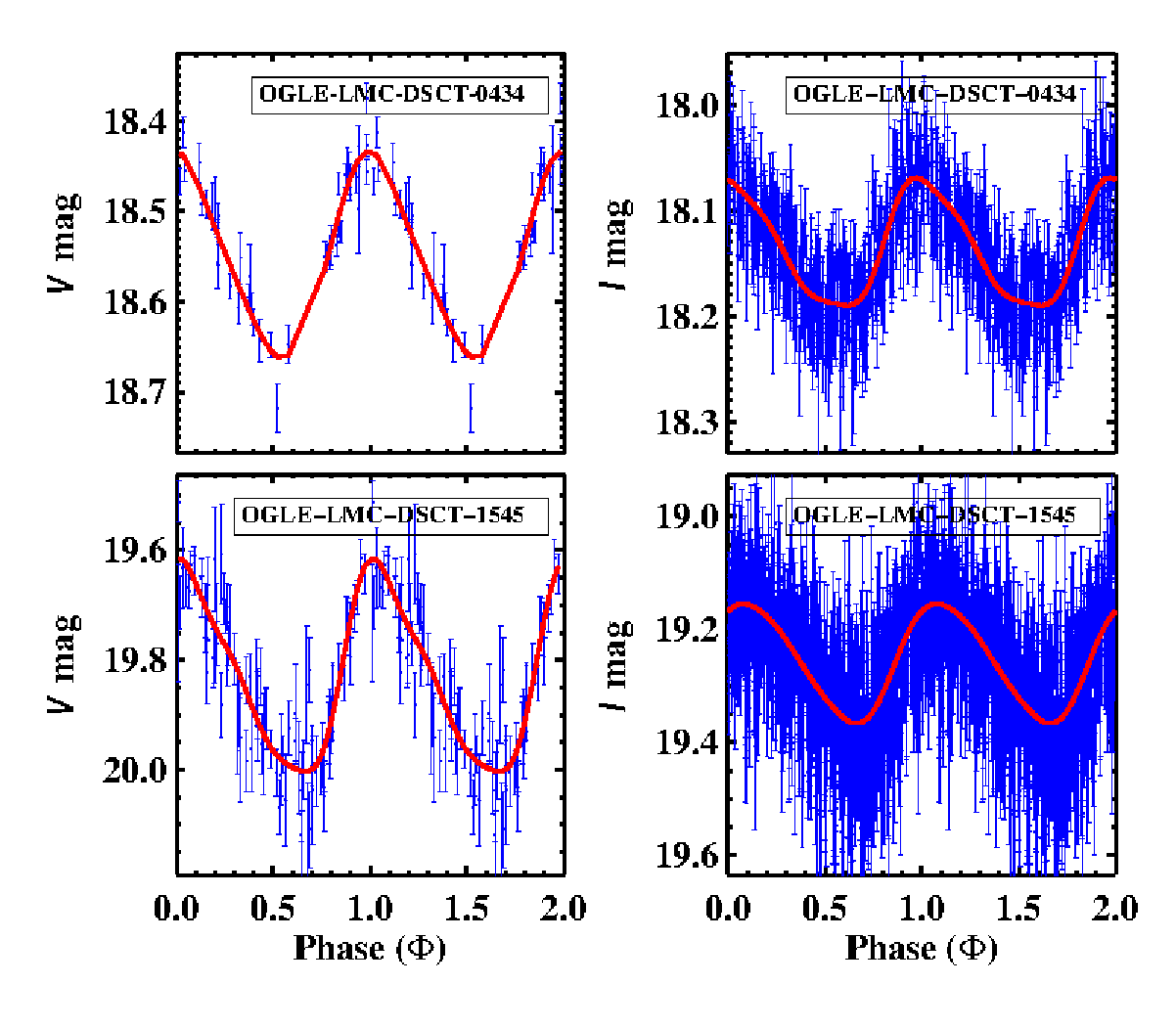}
\caption{Same as Fig.~\ref{fig:lc_blg} for LMC.}
\label{fig:lc_lmc}
\end{figure*}
\section{Selection and Cleaning of Data}
\label{sec:data}
The optical ($V,I$)-band photometric data and  light curves of $\delta$~Scuti 
stars belonging to the Galactic bulge are taken from the OGLE-IV database \citep{sosz21} and those for the LMC from  the OGLE-III database \citep{pole10}. At 
first, possible contaminant sources (foreground and background stars) are 
separated from the Galactic bulge stars and then removed using a 
colour-magnitude diagram (CMD). The CMD is constructed using the mean apparent 
$(V,I)$-band magnitudes as given in the database by closely following  \citet{piet15}. Secondly, the uncertain stars as listed in the `remarks.txt' file 
provided by  OGLE-IV (for the Galactic bulge) and OGLE-III (for the LMC) databases are removed. From this cleaned sample, finally, we choose well-sampled light curves of 
mono-periodic $\delta$~Scuti stars with more than $30$ data points for the analysis. The number of common stars in the final sample with complementary 
photometric data available in both the $I$- and $V$-bands for the Galactic bulge and  LMC are $3202$ and $1102$, respectively.
\section{Methods}
\label{sec:Methods} 
 The Fourier decomposition method is used to obtain the light curve parameters for the present study. Since the method can lead to numerical ringing when the raw data contains extreme outliers, the removal of these outliers from the raw light curve data is extremely important. The outliers  are removed from the raw light curves using the following condition \citep{leys13}: 
\begin{align*}
\frac{\left|m-\text{Median}(m)\right|}{\text{MAD}(m)} \ge & 3.0,
\end{align*}
where $m$ is the observed magnitude, $\text{MAD}$ represents the median absolute deviation. After carrying out extreme outlier removal steps on the raw light curves, the Fourier decomposition method is used to obtain the $V$- and $I$-band light 
curve parameters of the sample of $\delta$~Scuti stars.  The light curves are fitted with a Fourier sine series of the form  \citep{deb09} employing $3\sigma$ outlier clipping:
\begin{align}
\label{eq:Fourier}
m(t)=&A_{0}+\sum\limits_{i=1}^N A_{i} \sin[i\omega(t_{i}-t_{0})+\phi_{i}],
\end{align}
where $A_{0}$ represents mean magnitude, $\omega=\frac{2\pi}{P}$ is the angular frequency, and $P$ and $t_{i}$ denote the period 
of a star in days and the times of observations, respectively. The values of $P$ and $t_{0}$ are taken as given in the database. Here $t_{0}$ represents the 
epoch of maximum light for the  $V$- band which corresponds to 
maximum light at phase zero. The components $A_{i}$ and $\phi_{i}$ denote the $i$th order 
Fourier coefficients, and $N$ is the order of the fit.  The value of $N$ is 
taken to be $3$ and $4$ for LMC and Galactic bulge, respectively. Examples of Fourier fitted light curves are displayed in Fig.~\ref{fig:lc_blg} and Fig.~\ref{fig:lc_lmc}. Equation~(\ref{eq:Fourier}) has $2N+1$ unknown 
parameters. The light curves are phased using
\begin{align}
\label{eq:phase}
\Phi_{i}=&\frac{t_{i}-t_{0}}{P}-{\rm Int}\left(\frac{t_{i}-t_{0}}{P}\right).
\end{align}

From the Fourier fitted-light curve, the colours at maximum, mean and minimum 
light are obtained as follows:
\begin{align}
\label{eq:Colour}
(V-I)_{\rm max} &= V_{\rm max}-I_{\rm phmax} \\
(V-I)_{\rm mean} &= V_{\rm mean}-I_{\rm mean} \\
(V-I)_{\rm min} &= V_{\rm min}-I_{\rm phmin} 
\end{align}
where $V_{\rm max},V_{\rm mean}$ and $V_{\rm min}$ denote the maximum, mean 
and minimum magnitudes in the  $V$-band while $I_{\rm phmax},I_{\rm phmin}$ 
correspond to $I$-band magnitudes at the same phase of $V_{\rm max}$ and $V_{\rm min}$, respectively. Here $I_{\rm mean}$ is the mean $I$-band magnitude. 
This way of defining colours at maximum/minimum light allows for phase discrepancies between different bands.

To correct apparent magnitude values in the $V$- and $I$-band for the 
$\delta$~Scuti stars of the Galactic bulge, we use the \citet{nata13} 
extinction calculator, which is appropriate for the bulge.  For a given pair of 
$(l,b)$ values, the calculator returns the extinction value in the $I$-band $A_{I}$ 
as well as the colour excess $E(V-I)$. Using these two values, the 
extinction in the $V$-band $A_{V}$ is calculated. The \citet{nata13} 
calculator is based on  $E(J-K_{S})$ reddening values from \citet{gonz12}. 
Due to the non-standard nature of the reddening law  towards the Galactic bulge {\citep{popo00,udal03,nish09,nata13,mast13}}, the $A_{V}$ and $A_{I}$ values have also been calculated  adopting the standard \citet{card89} reddening law and $E(J-K_{S})$ values 
from \citet{gonz12} reddening map ($\frac{A_{J}}{A_{V}}=0.26;\frac{A_{I}}{A_{V}}=0.479$).

To correct for extinction in the LMC, $E(V-I)$ is obtained from the \citet{hasc11} 
reddening map which is converted into $E(B-V)$ values using 
$E(B-V)=\frac{E(V-I)}{1.38}$ \citep{tamm03}. The extinction values are 
obtained using \citet{card89} reddening law: $A_{V}=3.32E(B-V)$ and 
$A_{I}=1.94E(B-V)$ \citep{schl98}. The $E(V-I)$ values for the LMC are also 
obtained from the most accurate optical reddening map of \citet{skow21} for 
the Magellanic Clouds. The extinctions $A_{I}$ and $A_{V}$ are obtained 
using the relations $A_I=1.5E(V-I)$ and $A_{V}=2.5E(V-I)$ \citep{skow21}. 

Once the magnitudes and  the colours of the $\delta$~Scuti stars are corrected 
for extinction and reddening, we fit linear regression models to the PCAC 
relations employing an iterative $3\sigma$ outlier clipping. The results 
obtained from these analyses are discussed in the following section.
\section{Analysis and Results}
\label{sec:Results}
\subsection{PCAC relations}
\begin{figure*}
\vspace{0.014\linewidth}
\begin{tabular}{cc}
\vspace{+0.01\linewidth}
  \resizebox{0.45\linewidth}{!}{\includegraphics*{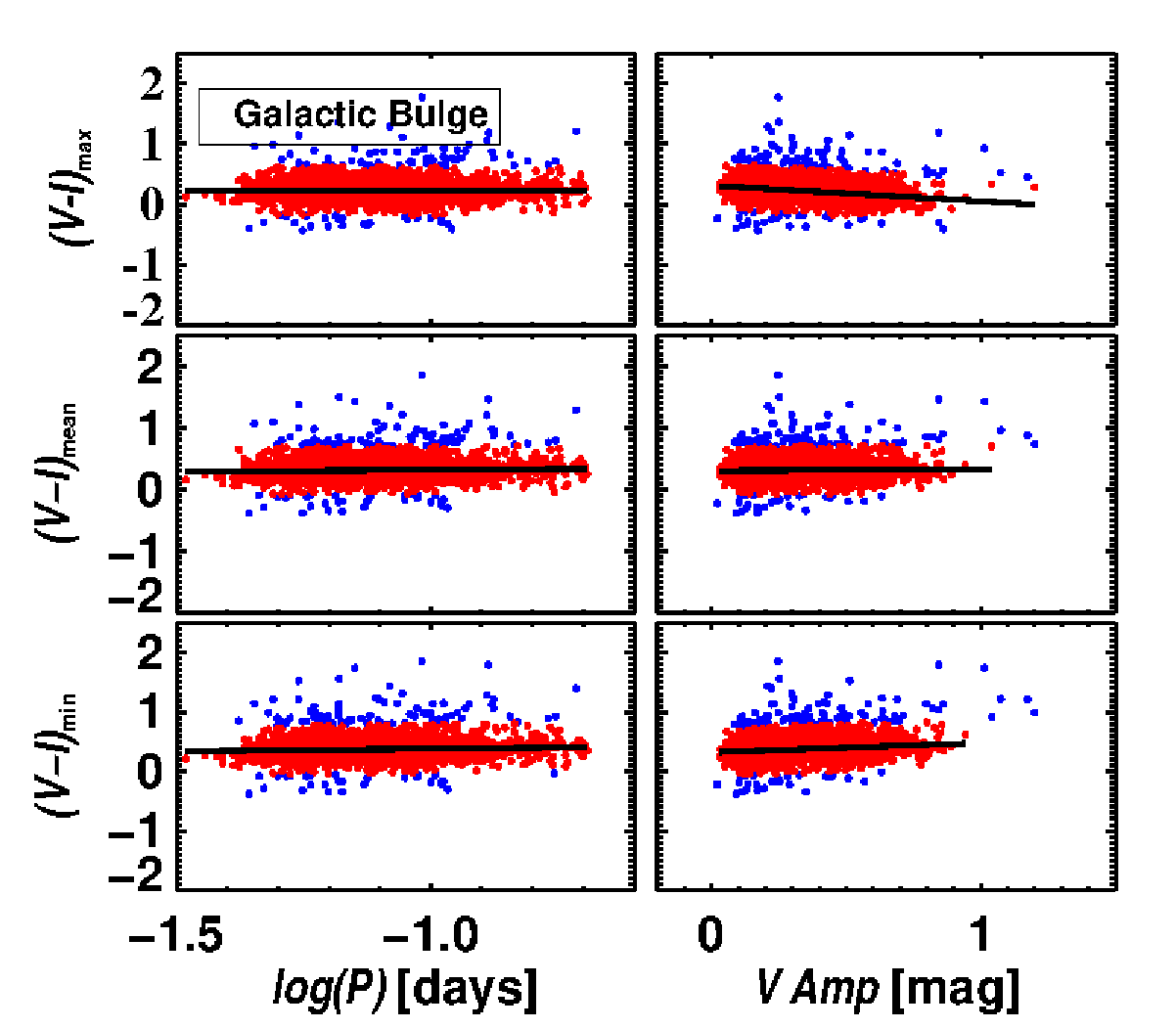}}&
  \resizebox{0.45\linewidth}{!}{\includegraphics*{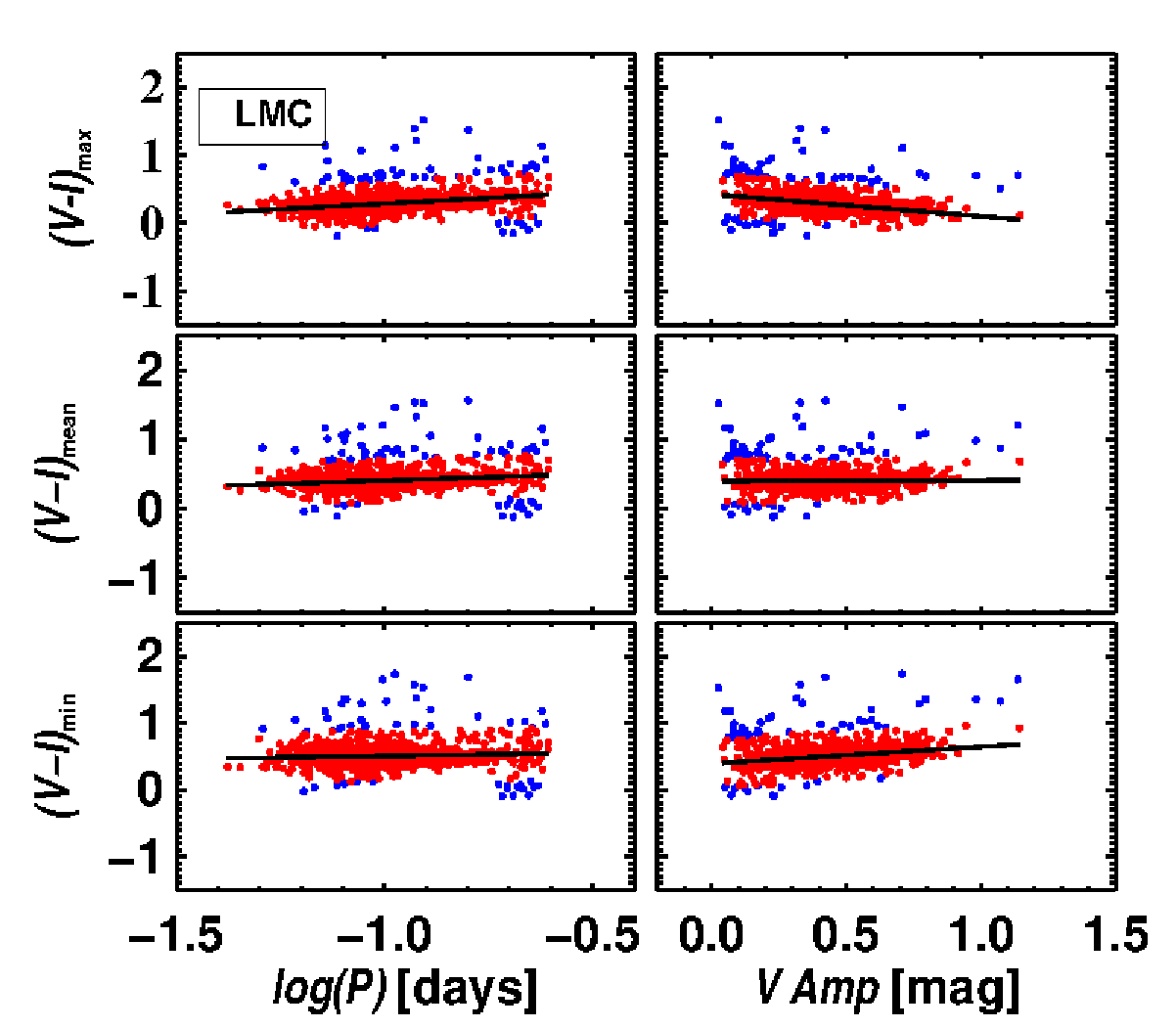}}\\
\vspace{-0.04\linewidth}
\end{tabular}
\caption{Left panel: PCAC diagram for Galactic bulge single-mode $\delta$~Scuti stars (with iterative 3$\sigma$ outliers clipping). The left/right sides are the PC/AC diagrams at maximum (top), mean (middle) and minimum (bottom).  The outliers are shown by the blue points. The solid lines represent the best fit to the plots. Right panel: Same for the LMC.}
\label{fig:pcac_blg}
\end{figure*}

\citet*{simo93} established that the PCAC relation at 
maximum and minimum light for Cepheids can be used to explore the radiation hydrodynamics of the star's outer envelope. They also explained the flat and 
sloped PC relations of Galactic Cepheids at maximum and minimum light, 
respectively, using the HIF-photosphere interaction theory. This theory has 
been verified based on the PCAC relations obtained using different types of 
pulsating stars \citep[][and references therein]{bhar14,ngeo17,das18,das20}. 
Here, we extend it to $\delta$~Scuti stars for the first time using the latest 
OGLE data available for the Galactic bulge and the LMC. The PCAC and PA plots for the Galactic bulge and the LMC $\delta$~Scuti stars are shown in Fig.~\ref{fig:pcac_blg} and ~\ref{fig:per_amp}, respectively. The results obtained from the PCAC and PA analyses  are summarized in Tables~\ref{table:result} and \ref{table:res_pa}, respectively.
\begin{figure*}
\vspace{0.014\linewidth}
\begin{tabular}{cc}
\vspace{+0.01\linewidth}
  \resizebox{0.45\linewidth}{!}{\includegraphics*{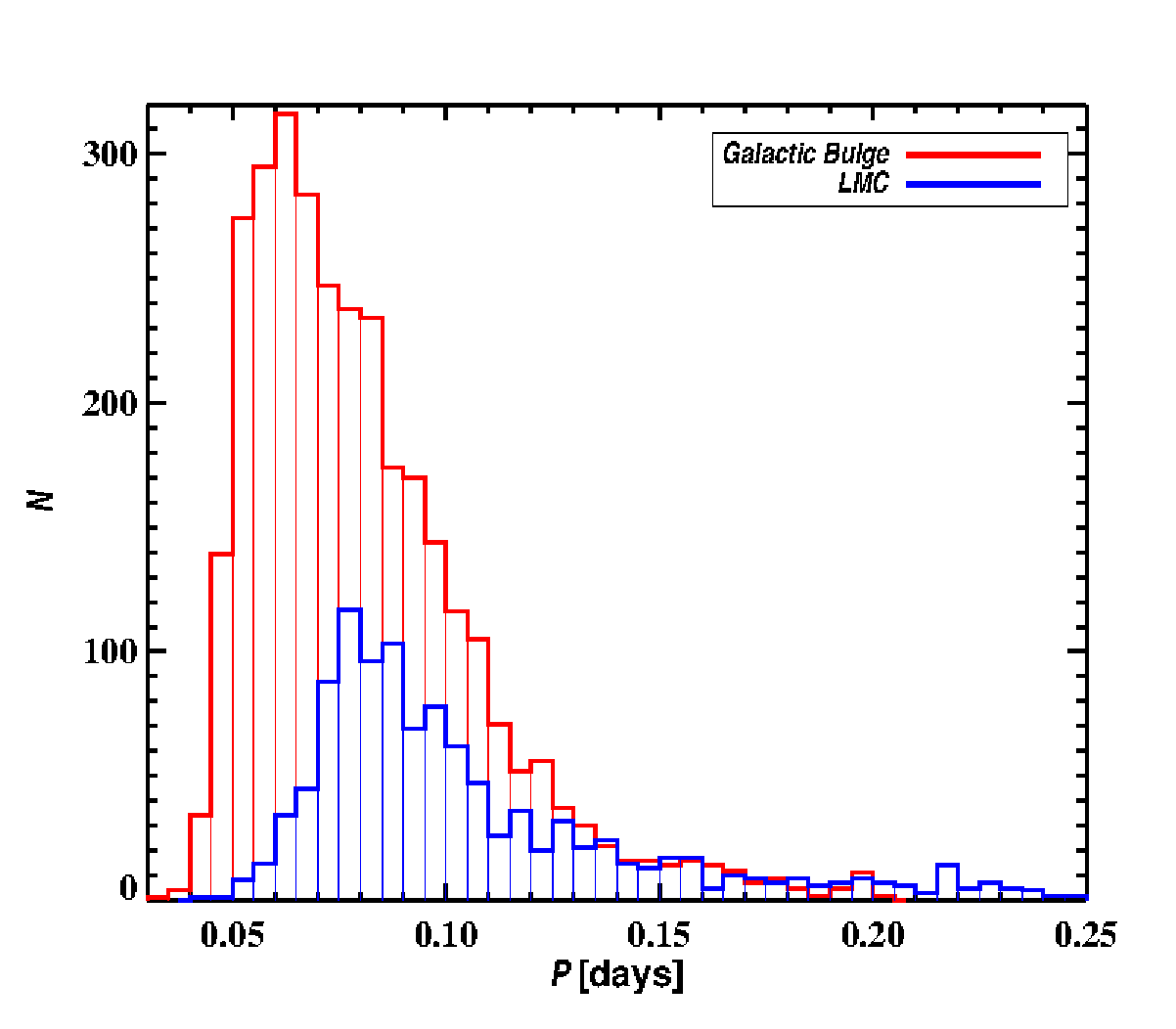}}&
  \resizebox{0.45\linewidth}{!}{\includegraphics*{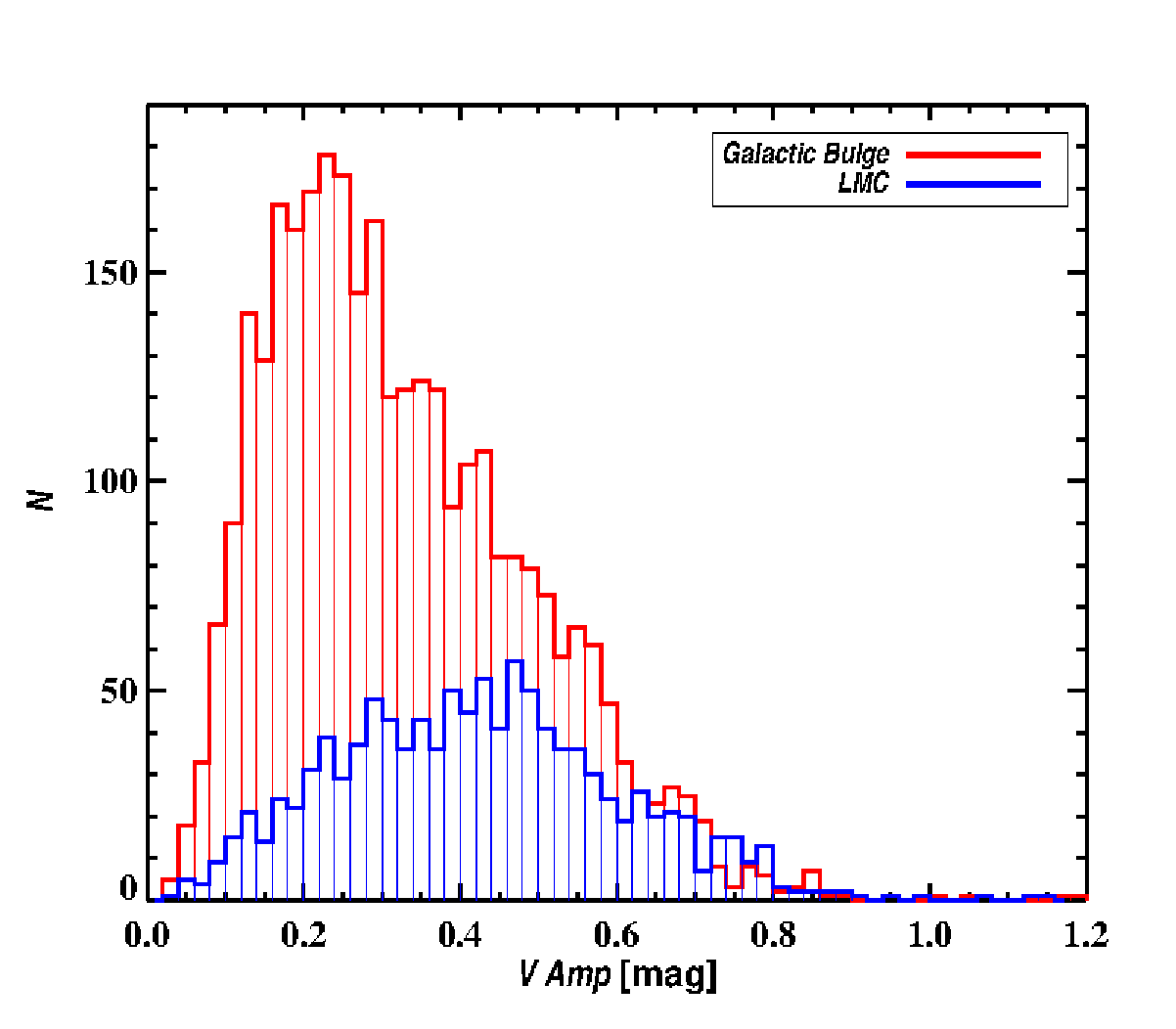}}\\
\vspace{-0.04\linewidth}
\end{tabular}
        \caption{Left panel: Histogram of period distributions of $\delta$~Scuti stars for both Galactic bulge (red) and LMC (blue). Right panel: Histogram of amplitude distributions of $\delta$~Scuti stars for both Galactic bulge (red) and LMC (blue).}
\label{fig:hist_pa}
\end{figure*}
\begin{figure*}
\vspace{0.014\linewidth}
\begin{tabular}{cc}
\vspace{+0.01\linewidth}
  \resizebox{0.45\linewidth}{!}{\includegraphics*{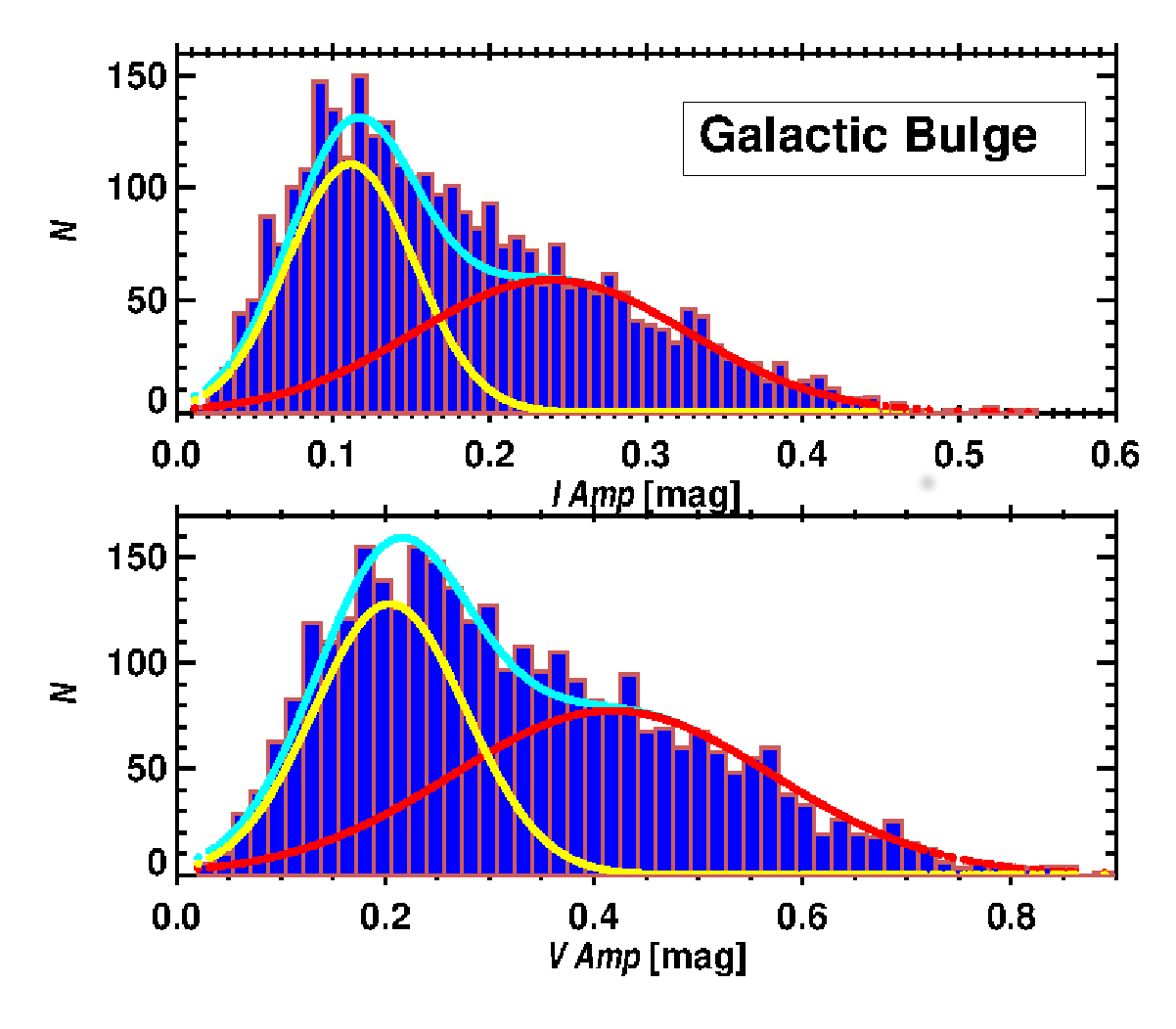}}&
  \resizebox{0.45\linewidth}{!}{\includegraphics*{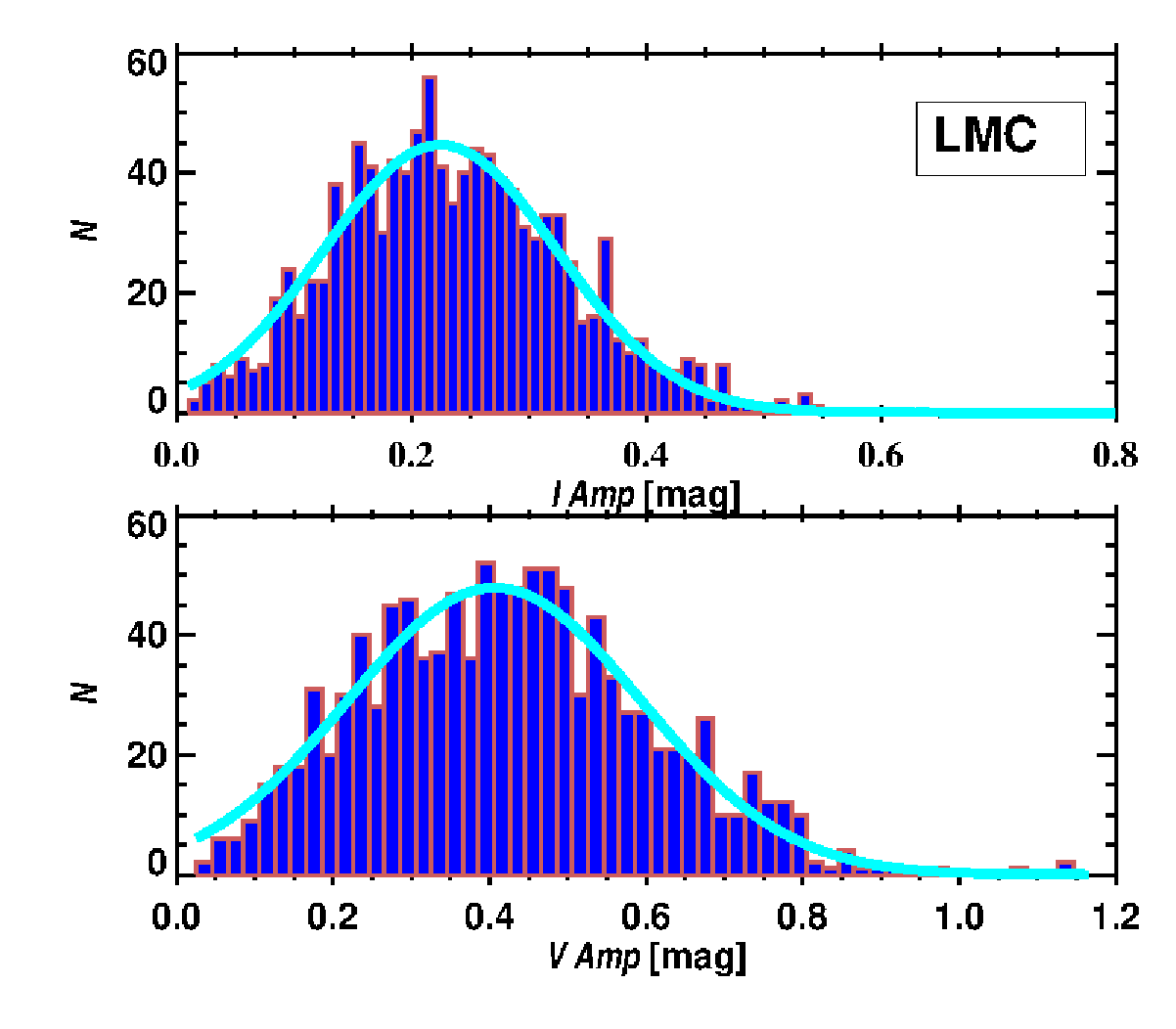}}\\
\vspace{-0.04\linewidth}
\end{tabular}
        \caption{The left panel displays amplitude histograms for Galactic $I$ (top figure) and $V$ (bottom figure). The cyan lines denote a two-component
        Gaussian fit to both amplitude distributions. The individual Gaussian fits are given by yellow and red lines, respectively. For $I$-band, the fits
        are centred at $0.111\pm 0.002$ (yellow-lower amplitude) and $0.240\pm 0.008$ (red-higher amplitude).  For the $V$-band, the corresponding values are
        centred at $0.204\pm 0.005$ and $0.418\pm  0.024$ for lower and higher amplitudes, respectively. The right panel presents the LMC amplitude histograms 
        for $I$ (top) and $V$ (bottom) bands, respectively. The histogram is consistent with a single Gaussian fit (cyan) centred at $0.224\pm 0.003$ and 
        $0.406\pm 0.006$ for the $I$- and $V$-band, respectively.
         }
\label{fig:gauss_amp}
\end{figure*}

\begin{figure*}
\vspace{0.014\linewidth}
\begin{tabular}{cc}
\vspace{+0.01\linewidth}
  \resizebox{0.45\linewidth}{!}{\includegraphics*{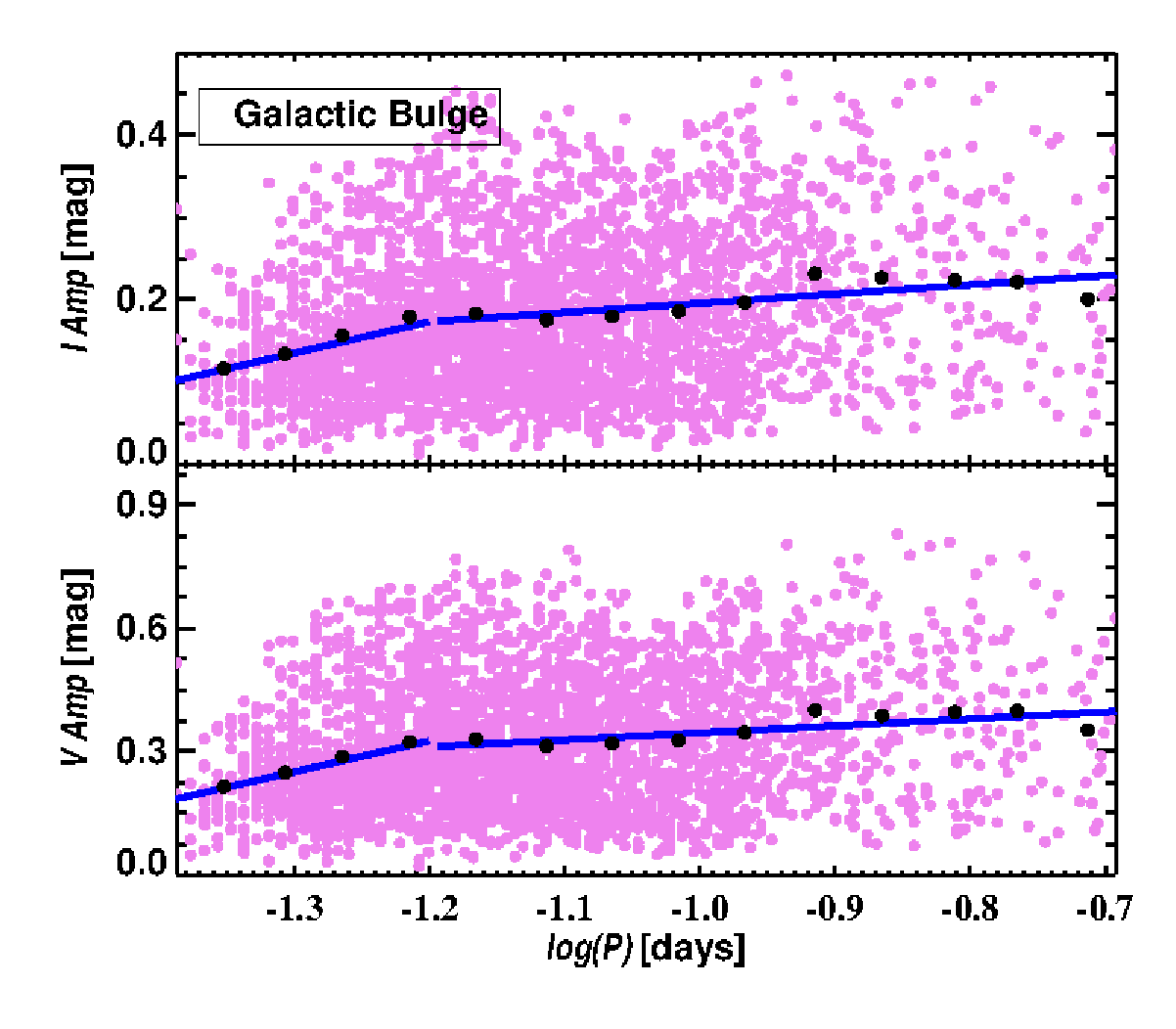}}&
  \resizebox{0.45\linewidth}{!}{\includegraphics*{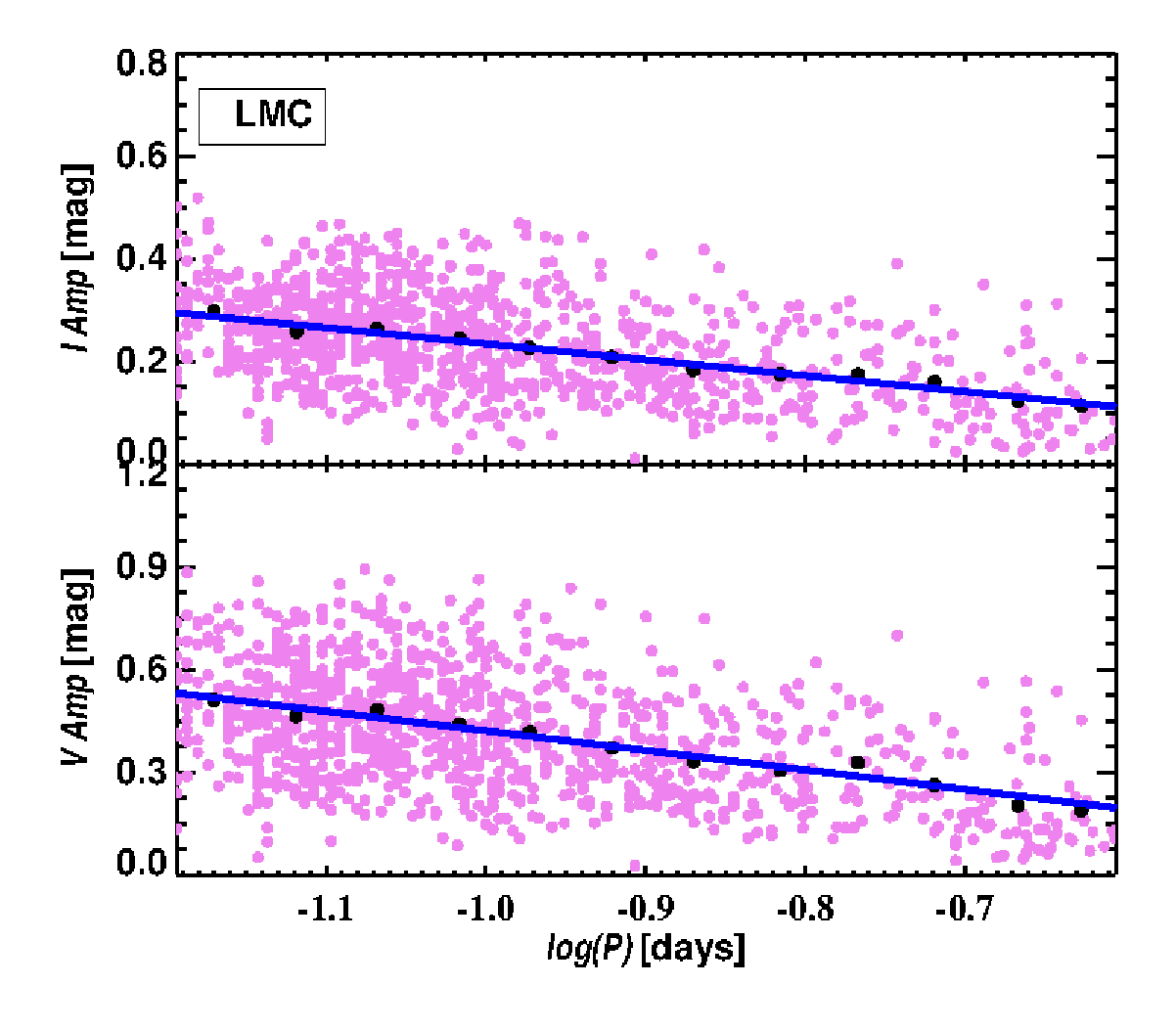}}\\
\vspace{-0.04\linewidth}
\end{tabular}
        \caption{Left panel: PA diagram for Galactic bulge single mode $\delta$~Scuti stars (with iterative 3$\sigma$ clipping). The blue lines represent the best-fit to the PA relations. There is a statistically significant break in the linear regression fit of Galactic bulge at $\log{P}=-1.2$. Binning of the data in every 0.05d in period (black dots) also shows the break in PA
        diagram at $\log{P}=-1.2$. Right panel: Same for LMC.}
\label{fig:per_amp}
\end{figure*}

\begin{table*}
\begin{threeparttable}[b]
	\centering
	\caption{ The slopes and intercepts of PCAC relation for Galactic bulge and LMC $\delta$~Scuti stars within $3\sigma$ uncertainty.}

	\label{table:result}
	\begin{tabular}{lcccccccccr} 
		\hline
		& Phase & Slope & Intercept & $\sigma$\tnote{a}&
		Nature of slope&$F$\tnote{b}&$P(F)$\tnote{c}&$R^{2}$\tnote{d}\\
		\hline
		\multicolumn{9}{|c|}{Galactic bulge \citep[Using reddening law from][]{nata13}} \\ \hline
		PC & Max & $0.002\pm0.017$ &$0.228\pm0.019$&0.132&Flat&$0.010$&$0.920$&$0.000$\\
		   & Mean& $0.054\pm0.016$ &$0.374\pm0.019$&0.126&Flat&$10.519$&$0.000$&$0.003$\\
		   & Min & $0.085\pm0.018$ &$0.467\pm0.020$&0.133&Flat&$22.696$&$0.000$&$0.007$\\
		AC & Max &$-0.257\pm0.014$ &$0.308\pm0.004$&0.122&Sloped&$352.500$&$0.000$&$0.103$\\
		   & Mean&$0.024\pm0.014$  &$0.304\pm0.005$&0.127&Flat&$3.862$&$0.082$&$0.001$\\
		   & Min &$0.161\pm0.014$  &$0.320\pm0.005$&0.132&Sloped&$117.331$&$0.000$&$0.037$\\ \hline
		   \multicolumn{9}{|c|}{Galactic bulge \citep[Using reddening law from][]{card89}} \\ \hline
		PC & Max & $-0.055\pm0.023$ &$0.053\pm0.026$&0.171&Flat&$8.257$&$0.004$&$0.002$\\
		   & Mean& $-0.020\pm0.022$ &$0.181\pm0.025$&0.167&Flat&$1.549$&$0.213$&$0.001$\\
		   & Min & $-0.000\pm0.023$ &$0.265\pm0.026$&0.168&Flat&$0.012$&$0.912$&$0.000$\\
		AC & Max &$-0.309\pm0.016$ &$0.216\pm0.006$&0.163&Sloped&$316.274$&$0.000$&$0.093$\\
		   & Mean&$-0.024\pm0.018$  &$0.212\pm0.006$&0.167&Flat&$2.844$&$0.092$&$0.001$\\
		   & Min &$0.122\pm0.019$  &$0.226\pm0.007$&0.168&Sloped&$32.186$&$0.000$&$0.010$\\ \hline
		   \multicolumn{9}{|c|}{LMC \citep[Using reddening map from][]{hasc11}} \\ \hline
		PC & Max & $0.328\pm0.022$ &$0.616\pm0.022$&0.098&Sloped&$223.824$&$0.000$&$0.177$\\
		   & Mean& $0.191\pm0.021$ &$0.598\pm0.021$&0.093&Sloped&$82.922$&$0.000$&$0.074$\\
		   & Min & $0.091\pm0.026$ &$0.599\pm0.026$&0.119&Flat&$11.868$&$0.001$&$0.011$\\
		AC & Max &$-0.325\pm0.016$ &$0.424\pm0.008$&0.091&Sloped&$377.734$&$0.000$&$0.268$\\
		   & Mean&$0.014\pm0.017$  &$0.396\pm0.008$&0.096&Flat&$0.733$&$0.392$&$0.001$\\
		   & Min &$0.252\pm0.020$  &$0.394\pm0.009$&0.112&Sloped&$152.250$&$0.000$&$0.127$\\ \hline
		   \multicolumn{9}{|c|}{LMC \citep[Using reddening map from][]{skow21}} \\ \hline
		PC & Max & $0.318\pm0.022$ &$0.572\pm0.022$&0.099&Sloped&$205.216$&$0.000$&$0.164$\\
		   & Mean& $0.186\pm0.021$ &$0.560\pm0.021$&0.092&Sloped&$81.158$&$0.000$&$0.072$\\
		   & Min & $0.087\pm0.026$ &$0.562\pm0.026$&0.119&Flat&$10.965$&$0.000$&$0.001$\\
		AC & Max &$-0.342\pm0.017$ &$0.400\pm0.008$&0.091&Sloped&$421.171$&$0.000$&$0.290$\\
		   & Mean&$0.002\pm0.017$  &$0.372\pm0.008$&0.095&Flat&$0.008$&$0.928$&$0.000$\\
		   & Min &$0.229\pm0.021$  &$0.373\pm0.009$&0.114&Sloped&$123.205$&$0.000$&$0.105$\\
		\hline
    
    \end{tabular}
    \begin{tablenotes}
    \item[a] Dispersions in the PCAC relations;
    \item[b] Value of $F$-statistic;
    \item[c] Probability of $F$-statistic;
    \item[d] Coefficient of variation for each linear regression.
  \end{tablenotes}
  \end{threeparttable}
\end{table*}

\begin{table*}
\begin{threeparttable}[b]
	\caption{The slopes and intercepts of PA relations for Galactic bulge 
and LMC $\delta$~Scuti stars within $3\sigma$ uncertainty.}
	\label{table:res_pa}
	\begin{tabular}{lccccccr} 
		\hline
		 & Slope & Intercept & $\sigma$\tnote{e}&$R^{2}$\tnote{f}& &\\
		\hline
		\multicolumn{6}{|c|}{Galactic bulge} \\ \hline
		 $V$ Band&$0.756\pm0.096$&$1.234\pm0.121$&$0.132$&$0.992$&$\rm{Period}<0.06~d$\\
		  & $0.170\pm0.032$&$0.515\pm0.034$&$0.158$&$0.987$&$\rm{Period}\ge0.06~d$\\ \hline
		 $I$ Band&$0.382\pm0.052$&$0.632\pm0.066$&$0.072$&$0.997$&$\rm{Period}<0.06~d$\\
		  & $0.112\pm0.019$&$0.308\pm0.020$&$0.094$&$0.994$&$\rm{Period}\ge0.06~d$\\ \hline 
		\multicolumn{6}{|c|}{LMC} \\ \hline
		   $V$ Band& $-0.569\pm0.034$ &$-0.149\pm0.034$&$0.152$&$0.393$&All period\\
		   $I$ Band& $-0.311\pm0.018$ &$-0.077\pm0.018$&$0.082$&$0.476$&All period\\
		\hline
	\end{tabular}
	\begin{tablenotes}
    \item[e] Dispersions in the PCAC relations;
    \item[f] Coefficient of variation for each linear regression.
  \end{tablenotes}
  \end{threeparttable}
\end{table*}

\begin{figure*}
\vspace{0.014\linewidth}
\begin{tabular}{cc}
\vspace{+0.01\linewidth}
  \resizebox{0.45\linewidth}{!}{\includegraphics*{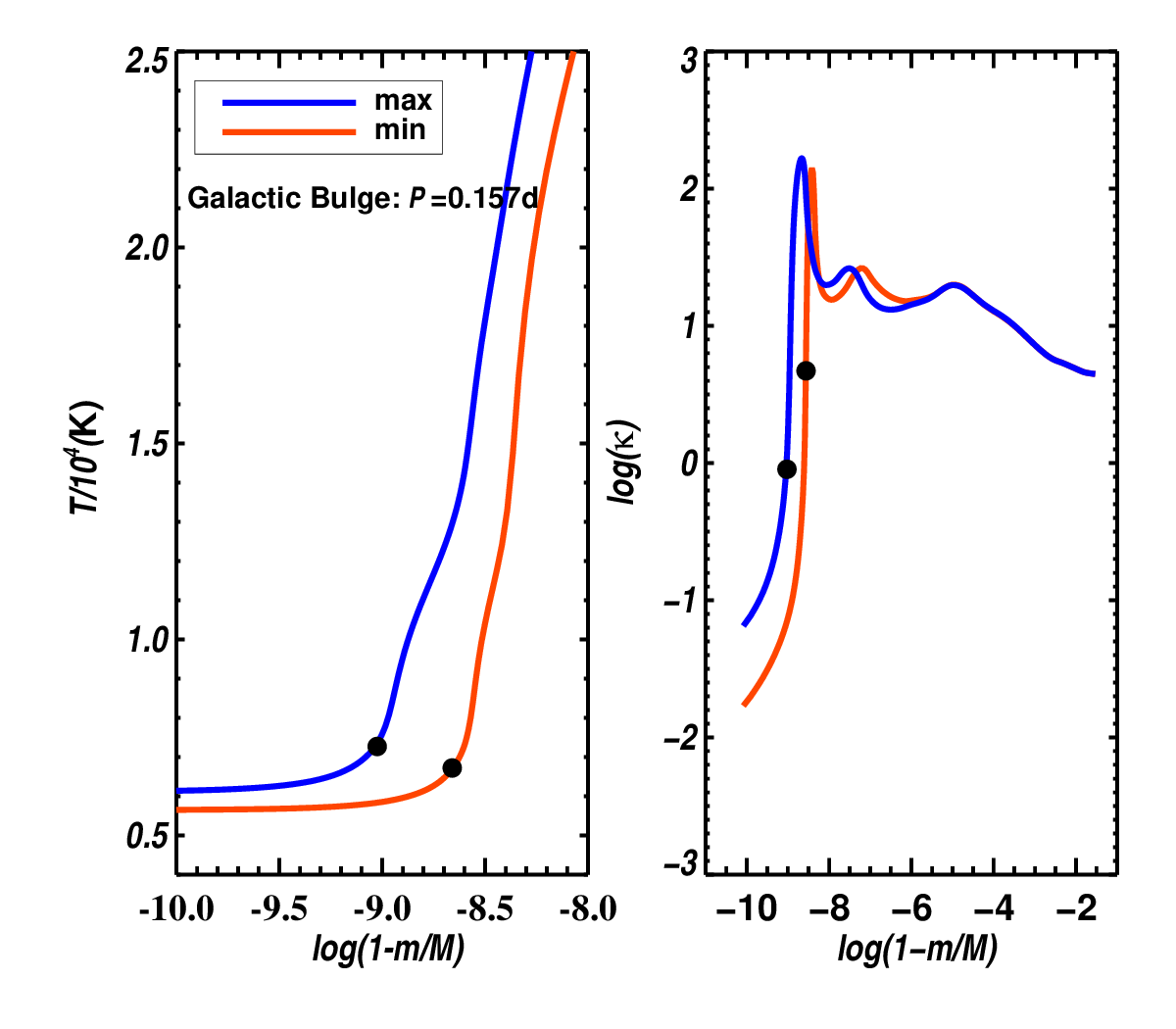}}&
  \resizebox{0.45\linewidth}{!}{\includegraphics*{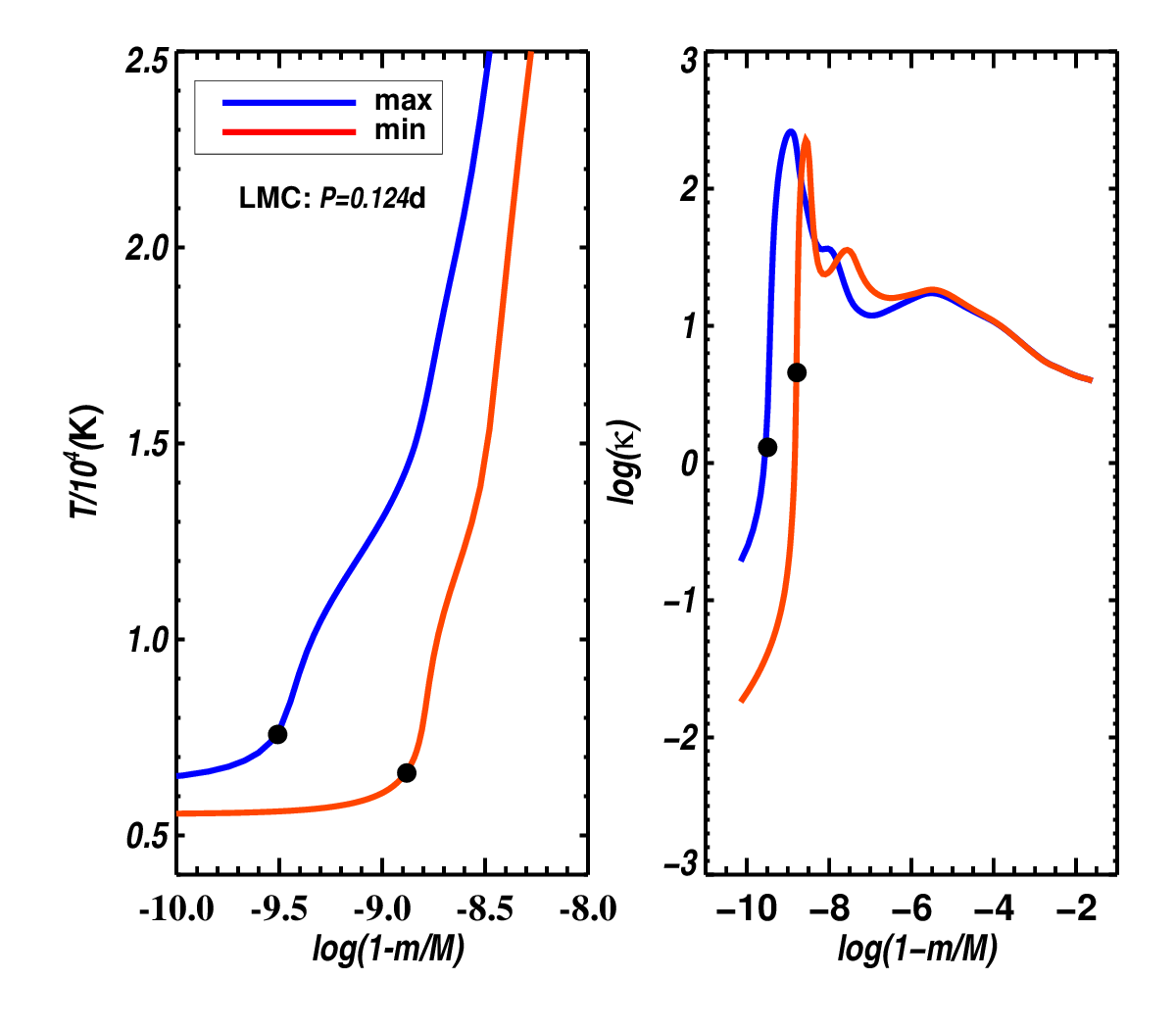}}\\
\vspace{-0.04\linewidth}
\end{tabular}
\caption{Theoretical temperature (left panel) and opacity profile (right panel) for a Galactic bulge (left figure)  and LMC (right figure)  $\delta$~Scuti star. The filled circles represent the location of the photosphere at $\tau=2/3$.}
\label{fig:mesa}
\end{figure*}

\subsubsection{Galactic Bulge $\delta$~Scuti}
The left panel of Fig.~\ref{fig:pcac_blg} shows the PCAC at 
maximum/mean/minimum light for the Galactic bulge $\delta$~Scuti stars. The parameters of the fitted relations are provided in Table~\ref{table:result}. The Galactic PC slopes at 
max/mean/min light are statistically close to zero. We observe significantly negative/positive AC slope at maximum/minimum light while the 
slopes of AC relations are flat at mean light.
\subsubsection{LMC $\delta$~Scuti}
The right panel of Fig.~\ref{fig:pcac_blg} shows the PCAC plots at 
maximum/mean/minimum light for the LMC $\delta$~Scuti stars. The solid line in 
black colour represents the fitted relations with the parameters as provided in 
Table~\ref{table:result}. We find  sloped PC relations at maximum/mean 
light and a flat relation at minimum light. Furthermore, the AC 
relation is sloped at maximum/minimum light while flat at mean light. The 
LMC $\delta$~Scuti stars are found to exhibit similar behaviour as displayed 
by the RRab stars when compared with the results obtained by \citet{das20}. These results are independent of reddening maps.

It is evident from Table~\ref{table:result} that the intercepts of PC relations at maximum/mean/minimum light of Galactic bulge stars have numerically smaller values than those for the LMC stars. A colour difference of $V-I\approx -0.20$ mag at mean light between the Galactic and LMC $\delta$ Scutis indicates that the Galactic $\delta$~Scutis are comparatively hotter than their LMC counterparts.

The dispersions in the PCAC relations of $\delta$~Scuti stars are similar to those of RR~Lyraes 
\citep{bhar14} and BL~Her stars \citep{das20}. To see whether the large dispersions are effecting the linear 
PCAC relations of $\delta$~Scuti stars or not, we made some further tests. 
We calculate the coefficient of variation $R^{2}$ for each linear regression as follows:
\begin{align}
R^{2}=1-\frac{\sum(y_{i}-\hat{y}_{i})}{\sum{(y_{i}-\overline{y})}}.
\end{align}
Here $y_{i}$ represents the observed reddening corrected colours of the stars, $\hat{y}_{i}$, the estimated values of $y_{i}$ 
form the linear relations and $\overline{y}$ is the mean value of $y_{i}$. The obtained values of $R^{2}$ are listed in Tables~\ref{table:result} and \ref{table:res_pa}.
The value of $R^{2}$ is close to zero for the Galactic bulge PC relations at all the phases, 
as expected. However, the intercepts of the 
Galactic PC relations at maximum and minimum light are statistically significantly different from each other. This is to be expected as the photosphere would be hotter at maximum light.  For the LMC PC relations, $R^{2}\approx0$ at minimum light, which supports the flat relation obtained, while $R^{2}>0$ at maximum/mean light supporting the sloped relation. An $F$-test following \citet{kanb04} furthermore demonstrates that the addition 
of the $\log{P}$ term in the PC relation makes a significant reduction in the error sum of squares term. 

Again, $R^{2}=0$ for the Galactic bulge AC relations at mean light and $R^{2}>0$ 
at maximum/minimum light supporting the nature of relation obtained in the analysis. Similarly, for 
LMC AC relations also, $R^2>0$ at maximum/minimum light and $R^2\approx0$ at mean light as expected. The $F$-test shows that the addition of the amplitude term in the AC relation also significantly reduces the error sum of squares. The $F$ statistic and corresponding 
$p(F)$, the probability of the $F$ statistic for each PCAC relation are mentioned in Table~\ref{table:result}.

\subsection{Period-Amplitude relation}

The character of the PCAC relations displayed in Fig.~\ref{fig:pcac_blg} may be understood in part by looking at Fig.~\ref{fig:hist_pa} which shows the histograms of the period and amplitude distribution. The left panel displays that the Galactic bulge $\delta$~Scuti stars are skewed towards shorter periods. The right panel implies that the LMC delta Scutis have higher amplitudes. Amplitude fluctuations are predominantly determined by temperature fluctuations. Thus, the lower/higher Galactic/LMC amplitudes are consistent with a flat/sloped PC relation at maximum light.

The left panel of Fig.~\ref{fig:gauss_amp} shows amplitude histograms for Galactic $I$ (top figure) and $V$ (bottom figure) bands. A two-component Gaussian fit (cyan colour) to the amplitude distribution of Galactic bulge $\delta$~Scuti stars fits the data very well. This clearly indicates two populations of $\delta$~Scutis: one with lower amplitude (yellow colour) and another with higher amplitude (red colour). For $I$-band, the fits  are centred at $0.111\pm 0.002$ (yellow-lower amplitude) and $0.240\pm 0.008$ (red-higher amplitude).  For the $V$-band, the corresponding values are centred at $0.204\pm 0.005$ and $0.418\pm  0.024$, respectively. The right panel presents the LMC amplitude histograms for $I$ (top) and $V$ (bottom) bands, respectively. The histogram is consistent with a single Gaussian fit (cyan) centred at $0.224\pm 0.003$ and $0.406\pm 0.006$ for the $I$- and $V$-band, respectively. Hence, the LMC $\delta$~Scutis are of higher amplitudes and  more evolved \citep{chan13}. From the intersection of the two Gaussians for the bulge, the lower/higher amplitude distribution dominates for $I$-band amplitude ($I$ Amp) less/greater than~$0.17$ mag and $V$-band amplitude ($V$ Amp) less/greater than $0.30$ mag. 

The OGLE-III LMC observed $\delta$ Scutis have mostly high amplitudes stars \citep{pole10}. They did not include many (more than $1400$) stars in their sample 
due to large photometric errors and/or small amplitudes. Hence, whilst the OGLE-IV Galactic $\delta$ Scuti sample may be complete, the OGLE-III LMC $\delta$ Scuti sample is likely to be missing a number of low amplitude stars. However, this provides an opportunity to compare a population of high amplitude $\delta$ Scutis in the LMC with a more mixed population in the Galactic bulge.

The $\delta$~Scuti PA plots for both Galaxy and LMC are given in Fig.~\ref{fig:per_amp}, and the results are summarized in 
Table~\ref{table:res_pa}. It is found that the Galactic bulge PA 
relation exhibits a break at period $0.06$~days, whereas no such break exists 
for the LMC $\delta$~Scuti stars. To test whether this break in the 
PA relation for the Galactic bulge $\delta$~Scutis is 
statistically significant, we use $F$-test \citep{kanb04}. From the 
$F$-test, the break in the PA relation at period 0.06~d is found 
to be statistically significant with an $F$-value equal to $8.662$ and $p(F)\approx0.0$.
Further binning of the data (black dots) in every $0.05$~d in period also supports the break in the PA relation at $\log{P}=-1.2$ as shown in Fig.~\ref{fig:per_amp}.

For both the Galactic and LMC PA relations, $R^{2}>0$ supporting the sloped 
relation. The $F$-test also confirms that the addition of $\log{P}$ in the PA 
relation significantly reduces the error sum of squares.

The PA relations display contrasting behaviour: for Galactic/LMC stars, the amplitude increases/decreases with period. The period-colour relation
indicates that Galactic $\delta$ Scuti stars are hotter than those in the LMC. Thus the Galactic blue edge $\delta$ Scuti instability strip is at a hotter 
temperature than that for the LMC. Thus, we have three regions: (i) the region mainly populated by Galactic $\delta$  Scutis and bordered by the Galactic blue edge and the LMC blue edge; (ii) the region populated by both Galactic and LMC $\delta$ Scutis bordered by the LMC blue edge and the Galactic red edge; (iii) the region mainly populated by LMC $\delta$ Scutis bordered by the Galactic red edge and the LMC red edge. As we go from region (i) to region (ii), the period and  amplitude are increasing (the positive slope in the Galactic PA diagram); from region (ii) to region (iii), the period is increasing and the amplitude decreasing. This explains the negative slope in the LMC PA diagram. However, further investigation is required to verify this postulate. 
\section{Discussion}
\label{sec:Discussion}
\begin{figure}
\includegraphics[width=0.45\textwidth,keepaspectratio]{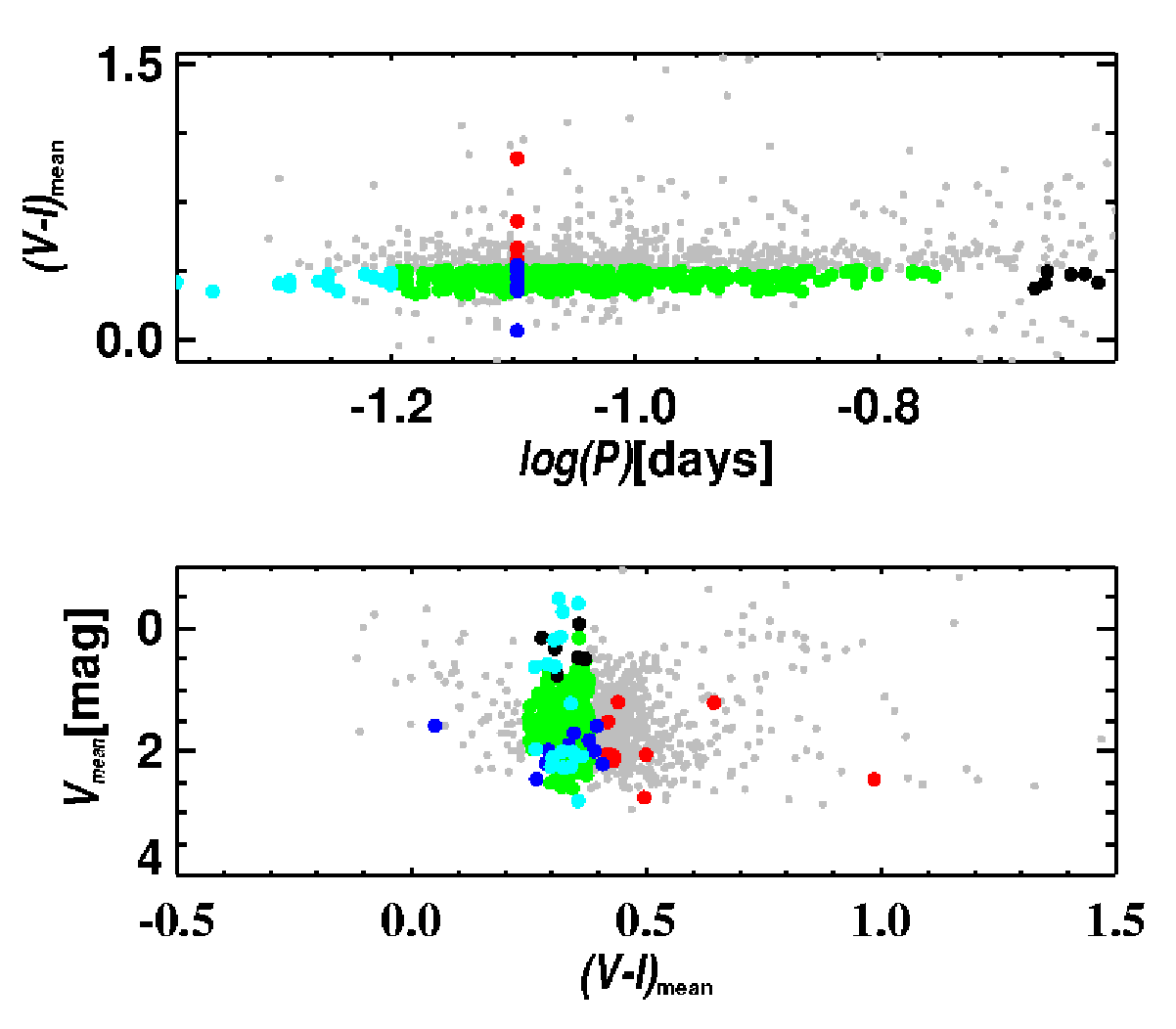}
\caption{Upper panel: PC diagram for LMC $\delta$~Scuti stars at mean light. Lower panel: CMD for the same. The blue and red points indicate the hotter and cooler stars at a fixed period ($P\approx0.08~{\rm d}$), respectively. The green 
points indicate the stars of similar colour, while the cyan and black dots indicate the 
stars of shorter and longer period but of same temperature.}
\label{fig:cmd_pc}
\end{figure}

The large dispersion in the PC relations may be explained in part using PC and CMD diagrams (Fig.~\ref{fig:cmd_pc}). In Fig.~\ref{fig:cmd_pc}. We consider the data around a fixed period, say $P\approx0.08~d,$ and apply simple arguments based on the period-mean density theorem. 
Because the blue stars in Fig.~\ref{fig:cmd_pc} have a  bluer colour, they are hotter and hence in order to have the same period, they need to have a higher $L/M$ ratio. Meanwhile, the red stars have a redder colour. They are cooler and hence in order to have the same period, they need to have a lower $L/M$ ratio. Thus, the blue and red stars on the PC diagram lie to the upper left and lower right, respectively, on a CMD diagram. 

Now, if we choose stars covering a narrow range of colour on the PC plane - the green stars in Fig.~\ref{fig:cmd_pc}. The black stars have longer period but the same temperature and hence must have a higher $L/M$ ratio, whilst the cyan stars have shorter period and the same temperature and hence must have a lower $L/M$ ratio. We suggest that this can explain some of the dispersion seen in the PC plots. Another source of dispersion in the PC relations may also 
be due to the amplitude variations observed in some $\delta$~Scuti stars \citep{bowm16}.

\subsection{A Possible Theoretical Explanation}
\label{sec:hif}
The HIF and photosphere move in the mass distribution of the star and are not 
necessarily co-moving during a pulsation cycle. When the HIF and stellar 
photosphere  are engaged (the stellar photosphere lies at the base of HIF), 
the temperature of the photosphere and hence the colour of the star are 
related to the temperature at which hydrogen ionizes. The temperature at which 
hydrogen ionizes is related to the properties of Saha ionization equilibrium: 
at low temperatures, the temperature at which hydrogen 
ionizes is largely independent of density and hence global stellar parameters. 
At higher temperatures, the temperature at which 
hydrogen ionizes is more dependent on the density and hence the stellar global 
parameters. Again, when the HIF and stellar photosphere are not engaged, then 
the photospheric temperature of the star has a stronger dependency on the 
stellar global parameters. As the colour is a measure of temperature and 
period is dependent on the stellar global parameters through the period-mean 
density relation, the relative location of HIF and photosphere at a 
particular pulsation phase will affect the corresponding PC relation \citep{das20}. 

The changes in the PC relations at maximum/minimum light and the corresponding 
behaviour in the AC relations were explained by \citet*{simo93} by applying 
Stefan-Boltzmann law at maximum and minimum light using the following equation:
\begin{align}
\label{eq:simon}
\log{T_{\rm max}}-\log{T_{\rm min}}=\frac{1}{10}(V_{\rm min}-V_{\rm max}),    
\end{align}
where $T_{\rm max}$ and $T_{\rm min}$ denote the effective photospheric 
temperature at maximum and minimum light, respectively. Equation~\ref{eq:simon} indicates that if $T_{\rm max}$ is independent or more weakly dependent on the pulsation period, then the changes in amplitude are related to the temperature at minimum 
light, leading to a correlation between the $V$-band amplitude and the observed colour at minimum light. Conversely, if 
$T_{\rm min}$ is independent or weakly dependent on the period, 
then the $V$-band amplitude and the observed colour will be correlated at maximum light.

 We have used the MESA-RSP version `MESA r15140' \citep{paxt10,paxt13,paxt15,paxt18,paxt19} which is based on the work of  \citet{smol08}  to compute two full-amplitude fundamental-mode models for $\delta$~Scuti stars with input parameters:
${Z=0.02}$, ${X=0.70}$, ${M=2.0~M_{\odot}}$, ${L=55~L_{\odot}}$, ${T=6950~K}$ (for Galactic bulge) 
and ${Z=0.008}$, ${X=0.736}$, ${M=1.6~M_{\odot}}$, ${L=25~L_{\odot}}$, ${T=6900~K}$ (for LMC). The model luminosities corresponding to the particular masses are
consistent with the evolutionary track for 
$\delta$~Scuti stars.  An example inlist used for computing the models is provided
in Appendix~\ref{app:inlist}. Mass fractions of hydrogen and metals are taken from the literature \citep{bono97,temp00}. The \citet{grev93} solar mixture is used to compute the models.  The two models presented in this work are computed using convection parameter set A as outlined in Table~4 of \citet{paxt19}. Detailed explanations of these free parameters can be found in \citet{smol08}.  For both models, the model envelope consists of $N=200$ Lagrangian mass cells, with $N_{\rm outer}=60$ cells between the anchor and the surface. Thus, the HIF is always covered by a large number of zones throughout the pulsation cycle.
The criteria used to check whether the model has reached full amplitude pulsation cycle are: the pulsation period $P$ computed on a cycle to cycle basis, the fractional growth of the kinetic energy per pulsation period $\Gamma$ and the amplitude of radius variation $\Delta R$ should not vary by more than $\sim 0.01$
over the last $\sim1000-$ cycles of the total integrations computed. As the growth rates of $\delta$~Scuti stars are very small, so it took $\sim 22000$ pulsation cycles for the Galactic model and $\sim 54000$ pulsation cycles for the LMC model to reach full amplitude stable pulsations for these models. Although MESA-RSP does not provide detailed atmosphere modelling and uses the diffusion 
approximation to determine the luminosity at different layers and bolometric corrections as a function of instantaneous temperature and effective gravity, previous work on PC relations for Cepheids and RR-Lyraes \citep[and references therein]{das20} has suggested that the diffusion approximation is adequate for the problems being discussed here.

Theoretical temperature and opacity profiles as function of mass distribution 
for each of the Galactic bulge and LMC models of $\delta$~Scuti stars are 
shown in Fig.~\ref{fig:mesa}. The temperature and the opacity profiles are 
obtained from the non-linear analysis using MESA-RSP \citep{paxt19}. The mass distribution is defined by the quantity $Q=\log{(1-\frac{m}{M})}$, where $m$ 
is the mass within radius $r$ and $M$ is total mass. The photosphere is 
defined as the zone with an optical depth $\tau=\frac{2}{3}$. These models 
suggest that the HIF and photosphere are engaged throughout the pulsation 
cycle in a manner similar to that of RR~Lyraes \citep{das20}.

For both the bulge and  LMC theoretical models, at minimum light, the HIF 
and the photosphere are always engaged at a temperature regime for which the 
ionization of hydrogen is somewhat weakly dependent or independent of temperature and hence the 
period. This explains the observed small slopes of PC relations  for both Galactic bulge 
and LMC $\delta$~Scuti stars at minimum light. On the other hand, the PC 
relation for the observed Galactic bulge $\delta$~Scuti stars at maximum light 
is flat,  while that for the LMC, there is a significant slope. 
Although the HIF and stellar-photosphere are engaged for both Galactic bulge
and LMC theoretical models (Fig.~\ref{fig:mesa}) at higher temperature, 
the observed bulge PC slope at maximum light is found to be close to zero. 
We note from these two non-linear models that the temperature of the photosphere at maximum light was higher for the LMC ($\sim 7576$~K) than the Galactic ($\sim7269$~K). The temperatures at minimum light were comparable. This suggests that one manifestation of higher amplitudes is that the HIF is driven further out in the mass distribution at maximum light. Even though Galactic $\delta$~Scuti stars are hotter, their temperature fluctuations are smaller than their LMC counterparts leading to smaller amplitudes. These smaller amplitudes may be due to different locations of the instability strip in the Galaxy and LMC. The smaller Galactic $\delta$~Scuti amplitudes, caused by smaller temperature fluctuations lead to a flatter PC relation at maximum light. Hence, we have good support for the HIF stellar photosphere theory as described in \citealt[and references therein]{simo93, das20}. 

We also note that there is some controversy over the pulsation mode of $\delta$~Scuti stars and in fact whether they may be pulsating in a number of modes and/or have a non-radial component \citep{netz22,chen22,khru22,netz22b}.  When we cross-checked our sample with the triple mode stars 
as listed in \citet{netz22}, we found $32$ such stars in our sample. Removal of these stars from the sample does not affect the nature of the PCAC relations. Furthermore, whether a star is singly periodic or has some combination of radial and even non-radial modes, the net result is some periodic relative motion of the HIF and photosphere in the mass distribution. It is this net motion between the HIF and photosphere that can determine the nature of the PC relation. The connection between PC and AC relations at two different phases does not contain any reference to mode of oscillation. Applying Stefan-Boltzmann's Law at maximum/minimum  \citep{simo93}, we get
${\frac{L_{\rm max}}{L_{\rm min}}} \approx {\frac{T_{\rm max}^4}{T_{\rm min}^4}},$
or
$\log L_{\rm max} - \log L_{\rm min} \approx \log T_{\rm max} - \log T_{\rm min},$
where $L_{\rm max}$ and $L_{\rm min}$ denote the maximum and minimum luminosity of a star, respectively and thereby determine the nature of the AC relation  at minimum/maximum light depending  on the behaviour of the PC relation at maximum/minimum light. In this case, a mixed population may indeed increase the dispersion of the PC or AC relation or lead to a different sloped relation. For example, first overtone RRc stars have a different sloped PC relation at minimum light as compared to fundamental mode RRab stars \citep[and references therein]{das20}. However, the fundamental relation given above is still valid as long as temperature variations are the primary cause of luminosity variations. Thus, to some extent, the HIF-photosphere theory outlined above is independent of this discussion. All that matters is the relative position of the HIF and stellar photosphere.

Investigating the PC relations of $\delta$~Scuti stars restricted purely to one mode will be a future project. \citet{das20} computed theoretical PC relations for three classes of variable stars using two different formulations of time dependent convection (corresponding to parameter sets A and D in \citet{paxt19,smol08}). They found broad agreement between models and theory under all of A and D. In future work, we plan to construct larger grids of models and study in greater detail how these theoretical PC relations may vary 
with the parameter sets A,B,C,D. 
\section{Summary and Conclusions}
\label{sec:Summary}
In this work, the $V$ and $I$-band light curves of $\delta$~Scuti stars 
belonging to the Galactic Bulge and  LMC are utilized from the publicly 
available  OGLE-IV and OGLE-III databases, respectively,  to study the 
PCAC relations for these stars. These relations are obtained at maximum, 
mean and minimum light  after applying iterative $3\sigma$ outliers removal  
and have been investigated for the first time using the largest available 
dataset. With this, we also extend the analysis of PCAC relation to much shorter periods ($\log{P}=-1.4$~d) which was done earlier down to $\log{P}=-0.5$~d \citep[Fig.~8 of][]{das20} adding $\delta$ Scuti stars to the classes of variable stars for which the HIF-stellar photosphere interaction theory has been successfully applied \citep{das20}. The results obtained in the present study are summarized as follows:
\begin{enumerate}
\item The observed PCAC relations for $\delta$~Scuti stars in the bulge and LMC are consistent with the HIF-stellar photosphere theory as outlined in \citet{das20} and references therein. 
\item The slopes of the PC relations for the Galactic bulge $\delta$~Scuti stars at maximum/mean/minimum light are flat and are statistically equal to zero. The AC relation is sloped at maximum/minimum light, while flat at mean light.
\item The PC relations for the LMC $\delta$~Scutis at maximum/mean light are sloped and flat at minimum light. The AC relations for LMC exhibit similar behaviour to that of the bulge, but the LMC AC relation has a relatively larger slope as compared to the bulge at maximum/minimum light.
\item The intercepts of PC relations at maximum/mean/minimum light of Galactic bulge stars have numerically smaller values as compared to those for the LMC stars. This indicates that
the bulge $\delta$~Scutis are comparatively hotter than the LMC $\delta$~Scutis.
\item Another important result obtained from this study is that the LMC short period $\delta$~Scutis have larger amplitude as compared to the bulge.
\item Two populations of $\delta$ Scuti in the Galactic bulge are evident from their amplitude distributions: one with lower amplitude and another with higher amplitude. Because the LMC is located at a distance of roughly $50~{\rm kpc}$, many stars in the OGLE-III $\delta$ Scuti database were not included due to large photometric uncertainty in small amplitude measurements. Hence, OGLE-III LMC $\delta$ Scuti samples might have missed a number of small amplitude $\delta$ Scutis. Due to this observational bias, OGLE-III observed $\delta$~Scutis have mostly high amplitudes.  
\item The PA relations are found to display contrasting behaviour: for Galactic/LMC stars, the amplitude increases/decreases with period. The period-colour relation shows that the Galactic blue edge $\delta$~Scuti instability strip is situated at a hotter temperature than that for the LMC. Considering three regions in the instability strip between the Galactic bulge blue edge and the LMC red edge, the observed positive/negative slope in the PA relation for Galactic bulge/LMC  has been explained. However, further investigation is required to verify the underlying postulate which we are planning to do in a future paper. 
\end{enumerate}
\section*{Acknowledgements}
The authors thank the reviewer for constructive comments and valuable suggestions which have greatly improved the presentation of the manuscript. We also acknowledge Prof. T. R. Bedding,  University of Sydney (Australia) for reading the first draft of the manuscript and providing many useful comments and suggestions. SD acknowledges Council of Scientific and Industrial Research (CSIR), Govt. of India, New Delhi for a financial support through the research grant 
``03(1425)/18/EMR-II''. MD thanks CSIR for providing the Junior Research Fellowship (JRF) through CSIR-NET under the project. SMK acknowledges the support of SUNY Oswego and Cotton University. S Das acknowledges the KKP-137523 `SeismoLab' \'Elvonal grant of the Hungarian Research, development and Innovation Office (NKFIH). KK is supported by  CSIR  senior research fellowship (SRF). AB acknowledges funding from the European
Union’s Horizon 2020 research and innovation programme under the Marie
Skłodowska-Curie grant agreement No. 886298. The authors acknowledge the use of High Performance Computing facility Pegasus at IUCAA, Pune and the following software used in this project:
MESA r15140 \citep{paxt10,paxt13,paxt15,paxt18,paxt19}. The paper makes use of the facility from \url{https://arxiv.org/archive/astro-ph}, NASA's Astrophysics Data System (ADS) and SIMBAD data bases.
\section*{Data Availability}
The data underlying this article are available at \url{http://ftp.astrouw.edu.pl/ogle/ogle4/OCVS/blg/dsct/}
and \url{http://ogle.astrouw.edu.pl/} for Galactic bulge and LMC, respectively. The derived data generated in this research will be shared on reasonable request to the corresponding authors.



\bibliographystyle{mnras}
\bibliography{dsct_pcac} 




\appendix
\clearpage
\section{MESA INLIST}
\label{app:inlist}
{\fontsize{6.2}{6pt}\selectfont
\begin{verbatim}
 &star_job

      show_log_description_at_start = .false.

      create_RSP_model = .true.

      save_model_when_terminate = .true.
      save_model_filename = 'final.mod'

      initial_zfracs = 2

      color_num_files=2
      color_file_names(2)='blackbody_johnson.dat'
      color_num_colors(2)=5

      set_initial_age = .true.
      initial_age = 0

      set_initial_model_number = .true.
      initial_model_number = 0

      profile_starting_model = .true.
       set_initial_cumulative_energy_error = .true.
      new_cumulative_energy_error = 0d0

/ ! end of star_job namelist
&eos
      use_FreeEOS = .true.
/

&kap
      kap_file_prefix = 'gn93'
      kap_lowT_prefix = 'lowT_fa05_gn93'
      kap_CO_prefix = 'gn93_co'
      Zbase = 0.008

! opacity controls
      cubic_interpolation_in_X = .false.
      cubic_interpolation_in_Z = .false.
      include_electron_conduction = .true.
      use_Zbase_for_Type1 = .true.
      use_Type2_opacities = .true.
      kap_Type2_full_off_X  =  0.71d0
      kap_Type2_full_on_X  =  0.70d0
      kap_Type2_full_off_dZ  =  1d-3
      kap_Type2_full_on_dZ  =  1d-2

/
&controls
 ! must set mass, Teff, L, X, and Z.
         RSP_mass = 1.6 ! (Msun)
         RSP_Teff = 6900 ! (K)
         RSP_L = 25 ! (Lsun)
         RSP_X = 0.736 ! hydrogen mass fraction
         RSP_Z = 0.008 ! metals mass fraction
        
         RSP_alfa = 1.2d0   ! mixing length; alfa = 0 gives a purely radiative model.
         RSP_alfac = 1.0d0  ! convective flux; Lc ~ RSP_alfac
         RSP_alfas = 1.0d0  ! turbulent source; Lc ~ 1/ALFAS; PII ~ RSP_alfas
         RSP_alfad = 1.0d0  ! turbulent dissipation; damp ~ RSP_alfad
         RSP_alfap = 0.0d0  ! turbulent pressure; Pt ~ alfap
         RSP_alfap = 0.0d0  ! turbulent pressure; Pt ~ alfap
         RSP_alfat = 0.0d0  ! turbulent flux; Lt ~ RSP_alfat; overshooting.
         RSP_alfam = 0.25d0 ! eddy viscosity; Chi & Eq ~ RSP_alfam
         RSP_gammar = 0.0d0 ! radiative losses; dampR ~ RSP_gammar    
        
         RSP_theta = 0.5d0  ! Pgas and Prad
         RSP_thetat = 0.5d0 ! Pturb
         RSP_thetae = 0.5d0 ! erad in terms using f_Edd
         RSP_thetaq = 1.0d0 ! avQ
         RSP_thetau = 1.0d0 ! Eq and Uq
         RSP_wtr = 0.6667d0 ! Lr
         RSP_wtc = 0.6667d0 ! Lc
         RSP_wtt = 0.6667d0 ! Lt
         RSP_gam = 1.0d0    ! Et src_snk
         
         ! controls for building the initial model
         RSP_nz = 200 ! total number of zones in initial model
         RSP_nz_outer = 60 ! number of zones in outer region of initial model
         RSP_T_anchor = 11d3 ! approx temperature at base of outer region
         RSP_T_inner = 2d6 ! T at inner boundary of initial model

         RSP_max_outer_dm_tries = 100 ! give up if fail to find outer dm in this many attempts
         RSP_max_inner_scale_tries = 100 ! give up if fail to find inner dm scale factor in this many attempts
         RSP_T_anchor_tolerance = 1d-8
         RSP_relax_initial_model = .true.
         RSP_relax_alfap_before_alfat = .true. ! else reverse the order
         RSP_relax_max_tries = 1000
         RSP_relax_dm_tolerance = 1d-6
         use_RSP_new_start_scheme = .false.
         RSP_kick_vsurf_km_per_sec = 1.0d0 
         RSP_fraction_1st_overtone = 0.1d0
         RSP_fraction_2nd_overtone = 0d0
        
        ! random initial velocity profile.  added to any kick from eigenvector.
         RSP_Avel = 0d0 ! kms. linear in mesh points from 0 at inner boundary to this at surface
         RSP_Arnd = 0d0 ! kms. random fluctuation at each mesh point.

         ! period controls
         RSP_target_steps_per_cycle = 600
         RSP_min_PERIOD_div_PERIODLIN = 0.5d0
         RSP_mode_for_setting_PERIODLIN = 0 
         RSP_default_PERIODLIN = 34560 

        ! when to stop
         RSP_max_num_periods =54000 ! ignore if < 0

         RSP_GREKM_avg_abs_frac_new = 0.1d0 ! fraction of new for updating avg at each cycle.
         ! timestep limiting
         RSP_initial_dt_factor = 1d-2 ! set initial timestep to this times linear period/target_steps_per_cycle
         RSP_v_div_cs_threshold_for_dt_limit = 0.8d0
         RSP_max_dt_times_min_dr_div_cs = 2d0 ! limit dt by this
         RSP_max_dt = -1 ! seconds
         RSP_report_limit_dt = .false.
         RSP_cq = 4.0d0 ! viscosity parameter (viscosity pressure proportional to cq)
         RSP_zsh = 0.1d0 ! "turn-on" compression in units of sound speed.
         RSP_Qvisc_linear = 0d0
         RSP_Qvisc_quadratic = 0d0
         RSP_use_Prad_for_Psurf = .false.
         RSP_use_atm_grey_with_kap_for_Psurf = .false.
         RSP_tau_surf_for_atm_grey_with_kap = 3d-3 ! for atm_grey_with_kap
         RSP_fixed_Psurf = .true.
         RSP_Psurf = 0d0 ! ignore if < 0.  else use as surface pressure.

         ! solver controls
         RSP_tol_max_corr = 1d-8
         RSP_tol_max_resid = 1d-6
         RSP_max_iters_per_try = 100
         RSP_max_retries_per_step = 8
         RSP_report_undercorrections = .false.
         RSP_nz_div_IBOTOM = 30d0 ! set IBOTOM = RSP_nz/RSP_nz_div_IBOTOM
         RSP_min_tau_for_turbulent_flux = 2d2
         ! rsp hooks
         use_other_RSP_linear_analysis = .false.
         use_other_RSP_build_model = .false.
         
         RSP_efl0 = 1.0d2
         RSP_nmodes = 3 
         RSP_trace_RSP_build_model = .false.

! output controls

      num_trace_history_values = 3
      trace_history_value_name(1) = 'rel_E_err'
      trace_history_value_name(2) = 'log_rel_run_E_err'
      trace_history_value_name(3) = 'rsp_GREKM_avg_abs'

      photo_interval = 1000
      profile_interval = 1
      history_interval = 1
      terminal_interval = 4000
      max_num_profile_models = -1
      log_directory='LOGS'
      photo_directory='photos'
/ ! end of controls namelist
&pgstar
/ ! end of pgstar namelist
\end{verbatim}
}


\bsp	
\label{lastpage}
\end{document}